\documentclass{main}
\usepackage{cite}
\usepackage{lineno}
\usepackage[export]{adjustbox}
\def\NIMA{{Nucl. Instrum. Methods} A}

\newcommand{\pt}{$p_{\rm T}$}

\def\be{\begin{equation}}
\def\ee{\end{equation}}
\def\bea{\begin{eqnarray}}
\def\eea{\end{eqnarray}}

\begin{document}

\title{\Large UPC physics with ALICE in Run 3}

\author{Anisa Khatun (for the ALICE Collaboration)}

\address{Department of Physics and Astronomy, The University of Kansas.\\ Lawrence, Kansas, 66045 USA
}

\maketitle 
\abstracts { 
The ALICE experiment has undergone a major detector upgrade for Run 3, expanding its detection capabilities for a wide variety of studies. The new continuous readout has significantly enhanced the physics potential for ultra-peripheral collision analyses. In this talk, we discussed some of the physics analyses that can be carried out in ultra-peripheral collisions using the Run 3 data and presented some of the first physics performance plots in both proton-proton and heavy-ion collisions.
}

\keywords{UPC, ALICE, Run 3, continuous readout}
%\linenumbers
\section{Introduction}
\begin{figure}[h!]
    \centering
\includegraphics[width=0.4\textwidth]{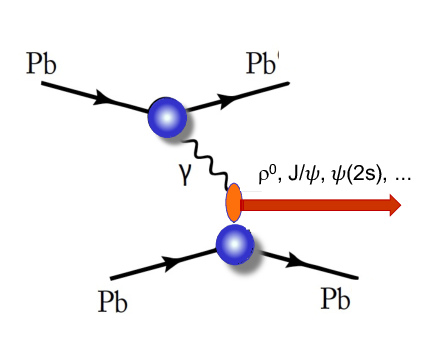}
   \includegraphics[scale=0.42]{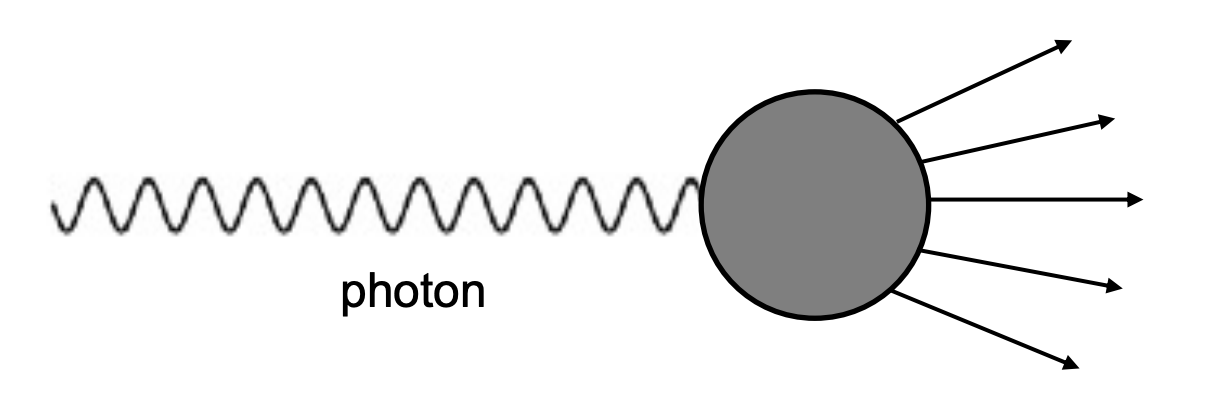}
    \caption{Typical exclusive vector meson production in UPC (left). The photon can also break up the target (right).}
    \label{fig:exclusiveinclusive}
\end{figure}

An ultra-peripheral collision (UPC) occurs when two relativistic heavy ions interact at very high energies with an impact parameter larger than the sum of the radii of the colliding nuclei. In a UPC, the two ions approach each other closely, but due to their high electric charge, they can interact electromagnetically without colliding nucleons. This leads to phenomena such as photonuclear interactions, where one or both ions emit photons that interact with the other. The hadronic process is suppressed in UPC due to the large impact parameter interaction. Either ion can emit a quasi-real photon, that can interact with a Pomeron and produce, for example, only a vector meson in the final state. This class of UPC interactions is known as the exclusive process (left Fig.~\ref{fig:exclusiveinclusive}). The exclusive process can be divided into two sub-classes: coherent process (photon interacts with whole nucleus) and incoherent (photon interacts with some of the nucleons). In the exclusive process, both the projectile and target remain intact. The photon can also break up the target as shown in the right Fig.~\ref{fig:exclusiveinclusive}. 

UPCs provide a great tool to probe the nucleus and nucleons. LHC can produce UPCs at luminosities and energies beyond what is achievable in any other collider. For example, the LHC can reach $\gamma p$ (photon-proton) energies ten times higher than those achieved at the Hadron-Electron Ring Accelerator (HERA). Consequently, in UPCs it is possible to probe lower Bjorken $- x$ values down to $10^{- 6}$ for the study of nuclear shadowing (and possibly gluon saturation) region of gluon distribution in the target nucleus~\cite{JandD}. ALICE has provided a rich variety of UPC physics studies with the data collected in the LHC Run 1 and Run 2~\cite{Alice2}. In the Run 3 program, ALICE continuous readout offers unique advantages for studying UPCs, including its good acceptance for both charged particles at low transverse momentum ($p_{\rm T}$) and excellent particle identification at midrapidity, which allows for precise characterization of the particles produced in UPCs \cite{Aliceupgrade}.

\section{The ALICE experiment in Run 3}\label{subsec:aliceexp}
In ALICE, one can study UPC physics at mid-, forward, and semi-forward (one track in the central barrel and one track in the muon arm) rapidity (see Fig.~\ref{eventdisplay}). At midrapidity, the track reconstruction and particle identification are done using data from the detectors: Inner Tracking System (ITS) \cite{ITSref}, Time Projection Chamber (TPC) \cite{TPCref}, Time of Flight (TOF) \cite{TOFref} and Transition Radiation Detector (TRD) \cite{TRDref}. At forward rapidity ALICE reconstructs muons using a muon spectrometer~\cite{Muonaremref}.
The interaction rate was 8 kHz for Pb–Pb and 100 kHz for pp collisions in Run 2. The ALICE experiment underwent a major upgrade to cope with an increased interaction rate of up to 50 kHz for Pb-Pb, 500 kHz for p-Pb and 1 MHz for pp collisions in continuous readout mode. A new Forward Interaction Trigger (FIT) replaced the V0 and T0 detectors and AD (used for trigger purposes in Run 2)~\cite{fitref}. The FIT detectors are placed on either side (A and C) of the interaction point, namely FT0A, FT0C, FV0A, FDDA, and FDDC. FIT provides veto on forward rapidity activity for UPC and diffractive physics studies and is also important for measuring the collision time and determining global collision parameters~\cite{fitref}. 

The upgraded ITS detector (ITS2) is a full silicon pixel detector and consists of 7 layers with an extended acceptance of pseudo-rapidity coverage $|\eta|<$ 1.22. ITS2 provides improved tracking efficiency, particle detection, and \pt\ resolution down to low \pt\ at 500 MeV/c. ITS2 is capable of providing a standalone reconstruction of tracks along with improved reconstruction of the primary and secondary vertices (originating from heavy-flavour hadrons)~\cite{ITS2ref}. TPC has been upgraded with a GEM-based readout to cope with the high interaction rate and also retains excellent particle identification~\cite{TPCupdate}. Both ITS2 and upgraded TPC are very important for UPC measurements of lower momentum particles at midrapidity ($|\eta|<$ 0.9). The readout for TOF, the Zero Degree Calorimeter (ZDC), and the muon arm have been improved. In particular, TOF has now an excellent time resolution of 56 $ps$ and improved particle identification which can be useful for UPC measurements for particles with relatively higher momentum~\cite{TOFupdate}. 

The MCH has been upgraded, and the muon trigger system has been replaced with the Muon Identifier (MID)~\cite{MIDref}. A new addition to the muon arm is the Muon Forward Tracker (MFT)~\cite{MFTref}. Especially, MFT extends tracking for charged particles to $|\eta| \sim$ 3.6 and also enables secondary vertexing for heavy-flavour measurements. Not only UPC quarkonia but also open heavy-flavour measurements are possible with MFT. The upgraded muon spectrometer with MFT can also be used to measure UPC low-mass vector mesons at forward rapidity thanks to reduced combinatorial background and improved mass resolution. The ZDC is used for UPC studies associated with neutron and proton emissions~\cite{ZDCref}.

\subsection{New common online-offline computing system}\label{subsec:fig}
ALICE now has a whole new software framework to process data samples more than 1000 times in pp and about 100 times in Pb-Pb collisions than the combined Run 1 and Run 2 samples at midrapidity. This software integrates online data taking using a new central trigger processor (CTP) and Data Quality Acquisition (DAQ)/Offline architecture as well as the analysis framework known as Online-Offline (O$^{2}$) software~\cite{pajipeter}. 

\begin{figure}[h!]
    \centering
    \includegraphics[width=0.85\textwidth]{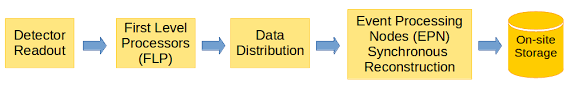}
    \caption{Schematic view of data-taking procedure.}
    \label{fig:dataflow}
\end{figure}

The ALICE Central Trigger Processor (CTP) was used in Run~2. The Local Trigger Unit (LTU) board for each
detector has provided the detector readout using the Trigger-Timing-Control (TTC). In Run~3, the ALICE detectors are self-triggered. The new CTP provides time stamps to synchronize data from different detectors, meaning no discrete events but a continuous data stream~\cite{trigger}. Software trigger is applied on the analysis level. The lack of any hardware trigger is referred to as 'trigger-less mode'. The data flow in the current Run 3 setup is shown in Fig.~\ref{fig:dataflow}.

\section{Progress on UPCs at ALICE}\label{sec:upcprogress}
\label{upceventrun3}
\begin{figure}[h!]
    %\centering
   \includegraphics[scale=0.42]{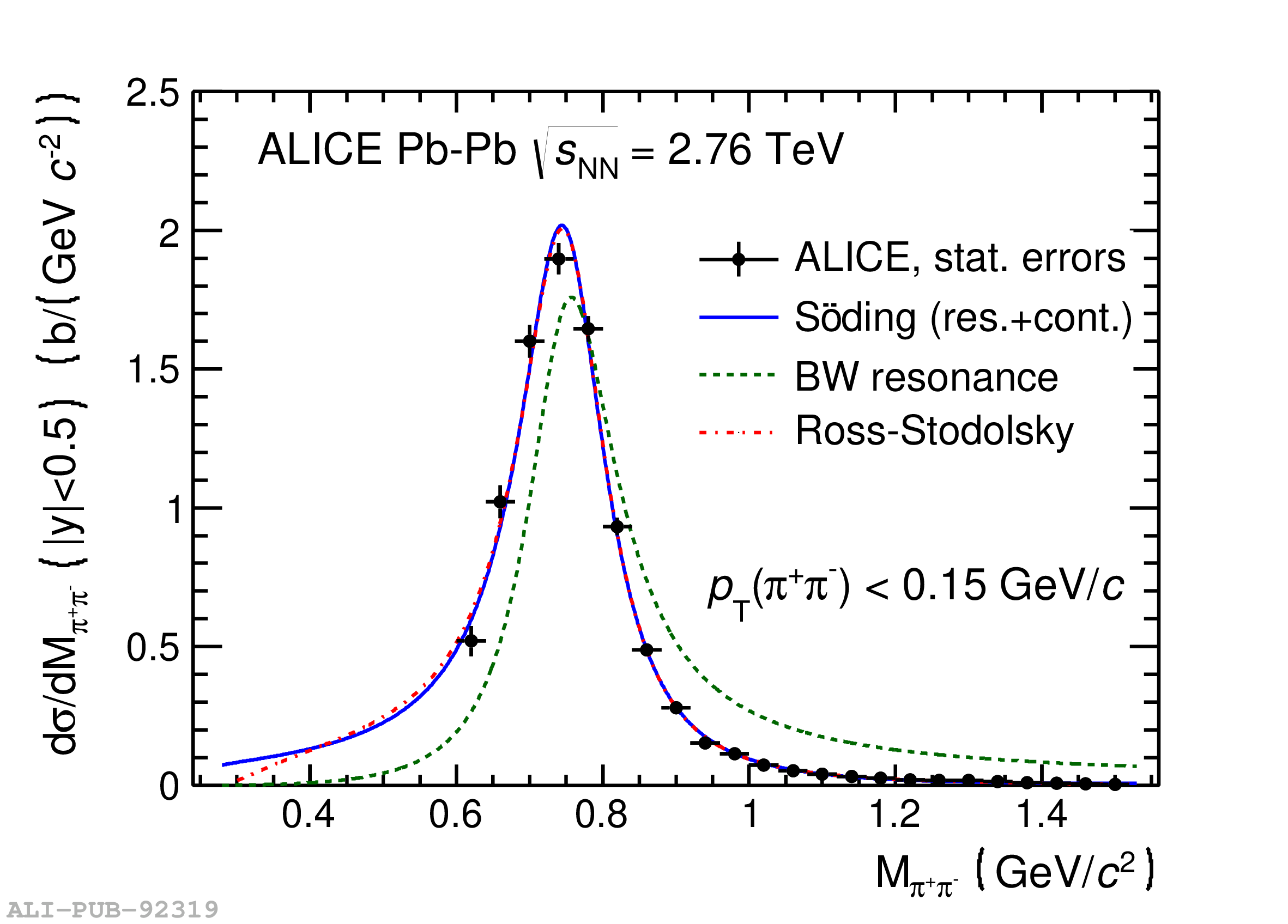}
\includegraphics[scale=0.40]{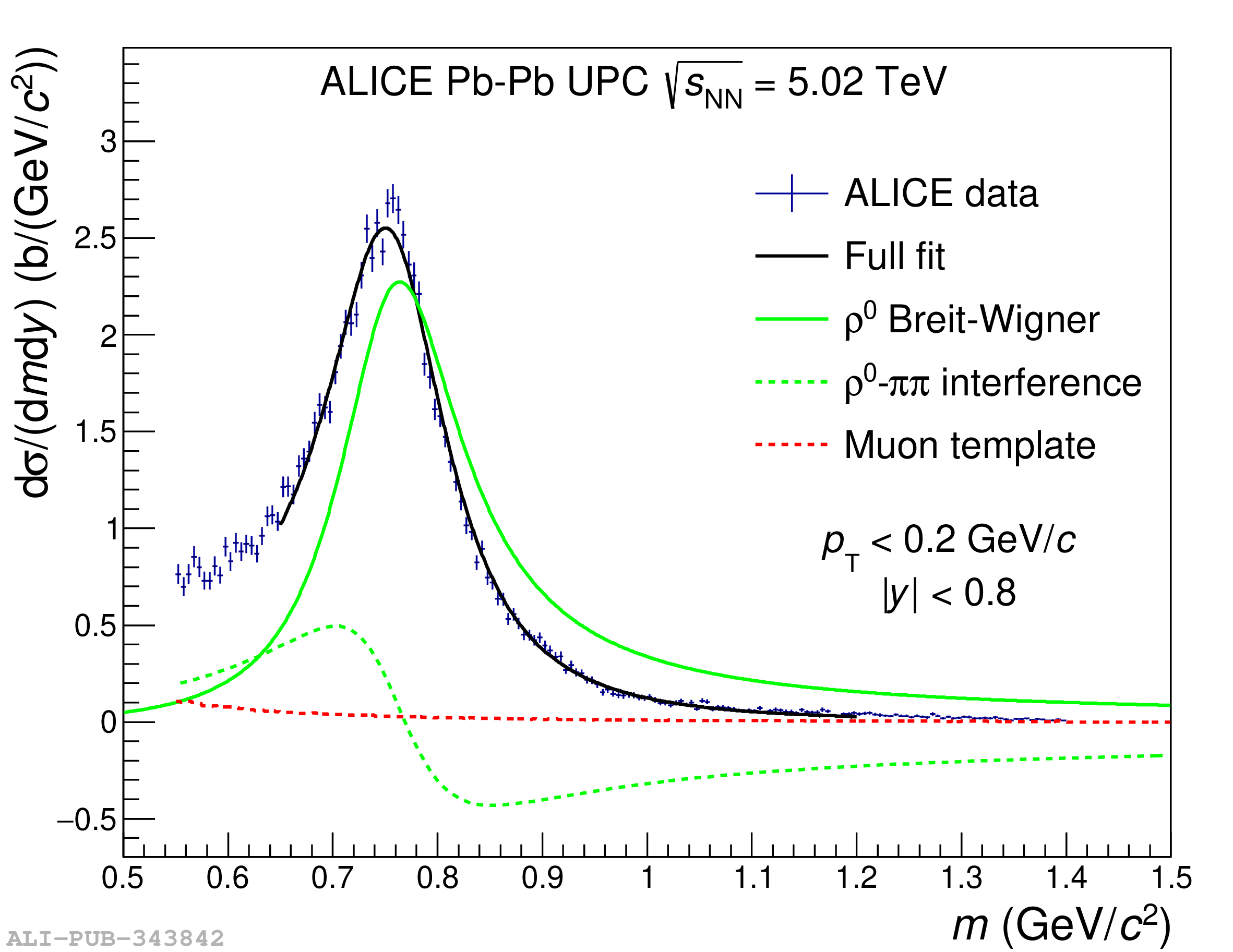}
    \caption{ a) Run 1 : $\sim$ 7k events \cite{rhorun1}. b) Run 2 : $\sim$ 60k events \cite{rhorun2}. }
    \label{fig:rhorun12}
\end{figure}

ALICE has come a long way with UPC physics from Run 1 to Run 2~\cite{Alice2}. If we take the example of coherent $\rho^{0}$ meson photoproduction, yield statistics had increased 10-fold during Run 2 at midrapidity (see Fig.~\ref{fig:rhorun12}). The collected integrated luminosity in Run 2 was 1 $\rm nb^{-1}$, and the expected luminosity to reach during Run 3 and Run 4 is 13 $ \rm nb^{-1}$ in total. A significant increase in statistics in Run 3 is possible with continuous readout~\cite{yellowreport}. 

\subsection{Datasets available for UPC in Run 3}
 %\label(upcrun3datataking)
 First Pb-Pb data taking with 2 kHz hadronic interaction rate was recorded on $26^{\rm th}$ September 2023. Data are taken successfully with 45 kHz Hadronic Interaction Rate. The total collected integrated luminosity in 2023 is about 1.5 $\rm nb^{-1}$ for UPC events. The data sample for minimum bias events collected in Run 3 at the end of 2023 is 40 times larger than Run 1 and Run 2 combined. The value 1.5 $\rm nb^{-1}$ is 3000 times larger than integrated luminosity used in $\rho^{0}$ analysis of Run 2 data. About 29 $\rm pb^{-1}$ of integrated luminosity was recorded in pp collisions at $\sqrt{s}$ = 13.6 TeV in 2022 and 2023. The collected luminosity of total triggered events was $\sim$ 8 $\rm pb^{-1}$ in Run 2 for pp collisions. The data in pp collisions are important for studying Central Exclusive Production (CEP), which is discussed in Section \ref{cepsec}. In Run 2, the study of CEP events was not possible since the collected data with high multiplicity in V0 and MUON, Electromagnetic Calorimeter and TRD triggers were most likely to exclude all CEP events in pp collisions.
 
\iffalse
\begin{figure}[h!]
    %\centering
    \includegraphics[scale=0.41]{r_integrated_luminosity_2023_3.pdf}
    \includegraphics[scale=0.37]{r_integrated_luminosity.png}
    \caption{Recorded Luminosity in Run 3 in Pb-Pb (left) and pp (right) collisions.}
    \label{fig:enter-label}
\end{figure}
\fi
\subsection{UPC event selection}
\begin{figure}[h!]
    %\centering
    \includegraphics[width=0.48\textwidth]{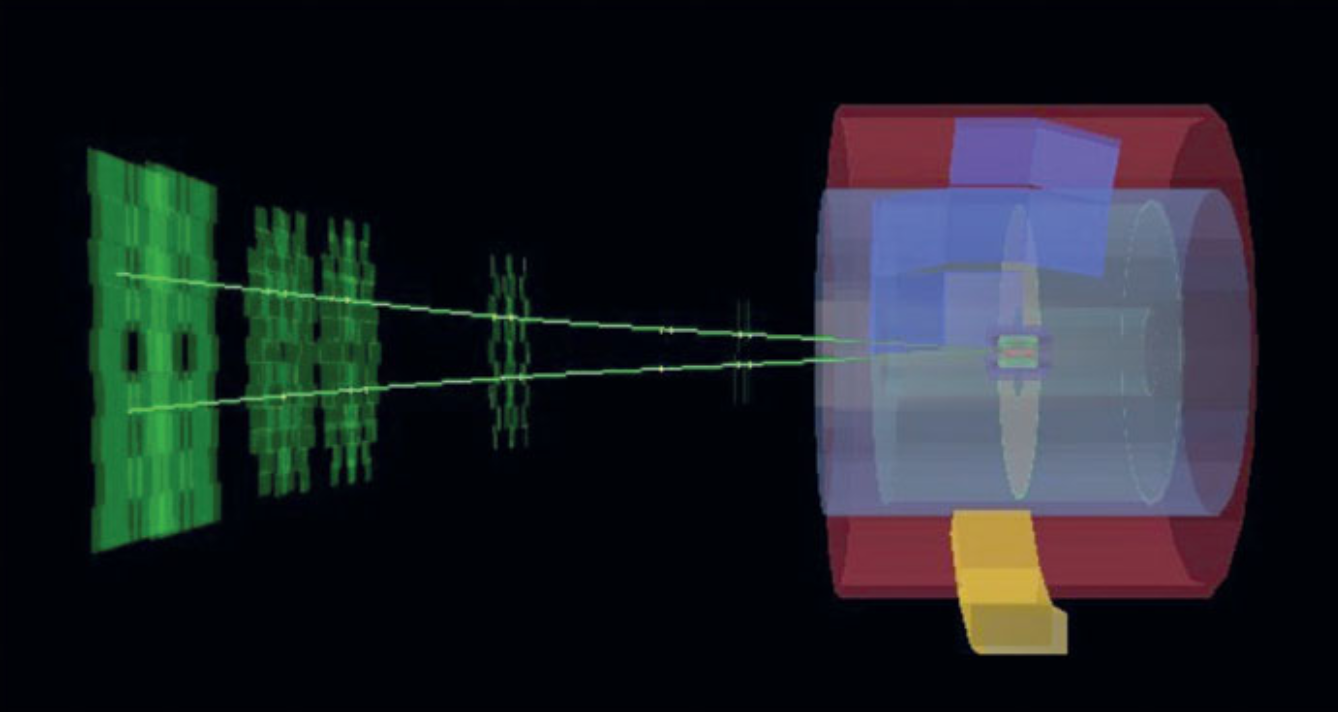}
    \includegraphics[width=0.45\textwidth]{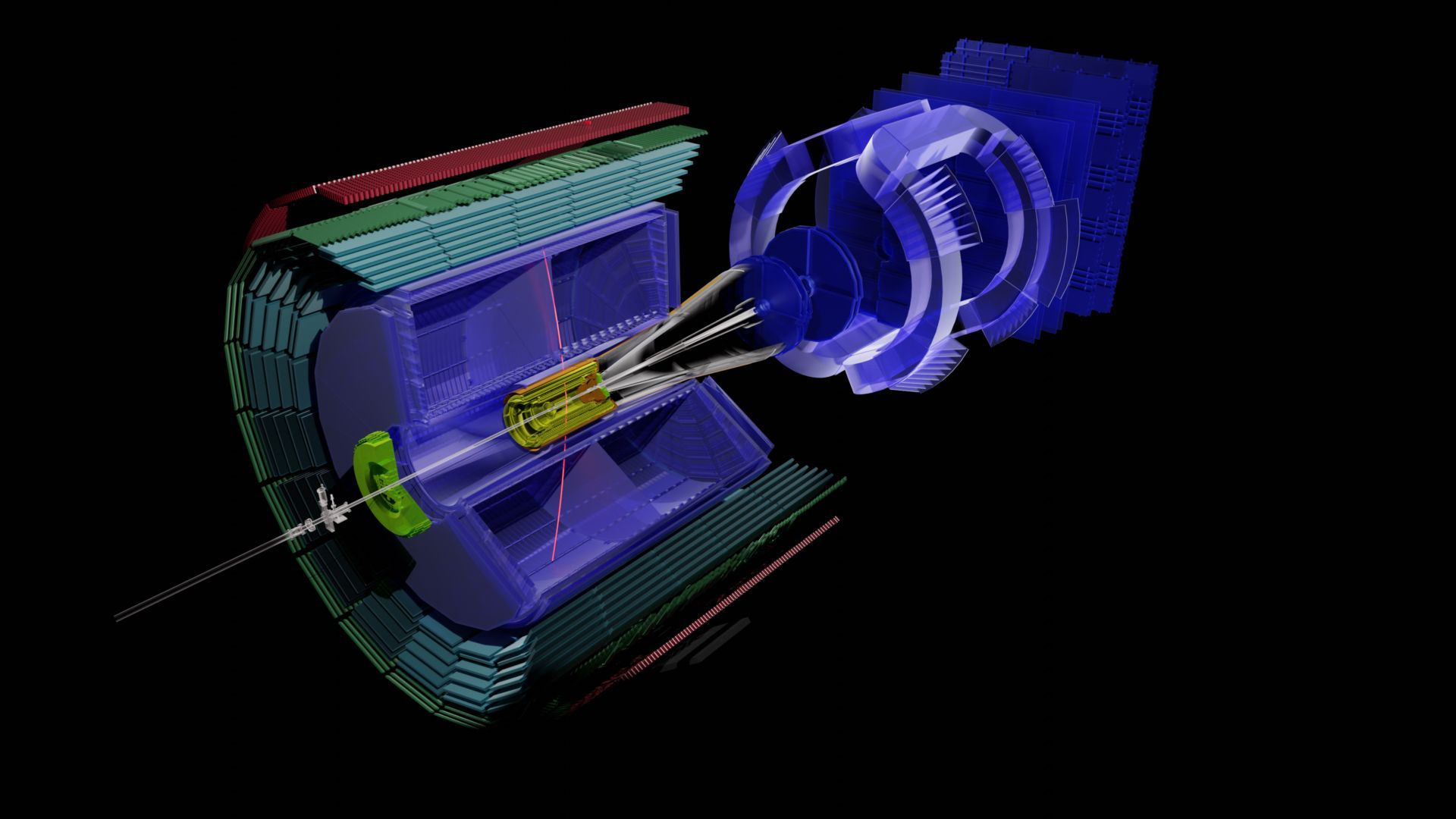}
    \caption{ UPC event in Run 2 at forward rapidity (left) and UPC event in Run 3 at midrapidity (right). }
    \label{eventdisplay}
\end{figure}

As in Run 2, selecting an exclusive vector meson requires two opposite sign tracks with no signal in the FIT detectors (see Fig.~\ref{eventdisplay}). The coherent ($p^{\rm VM}_{\rm T} \sim 1/R_{\rm Pb}$ = 50 MeV) and incoherent ($p^{\rm VM}_{\rm T} \sim 1/R_{\rm p}$ = 400 MeV) can be separated by selecting the $p_{T}$ scale of the system. Further, a veto can be applied on the ZDC signal for truly exclusive events or associated neutron emission processes can be studied using the ZDC signal. The ALICE trigger-less readout offers the possibility to veto signals coming from the individual A-side and C-side FIT subdetectors and performing a selection of inclusive and semi-inclusive UPC events.

\subsection{Ongoing UPC activity in Run 3}
\begin{figure}[h!]
    %\centering
    \includegraphics[width=0.48\textwidth]{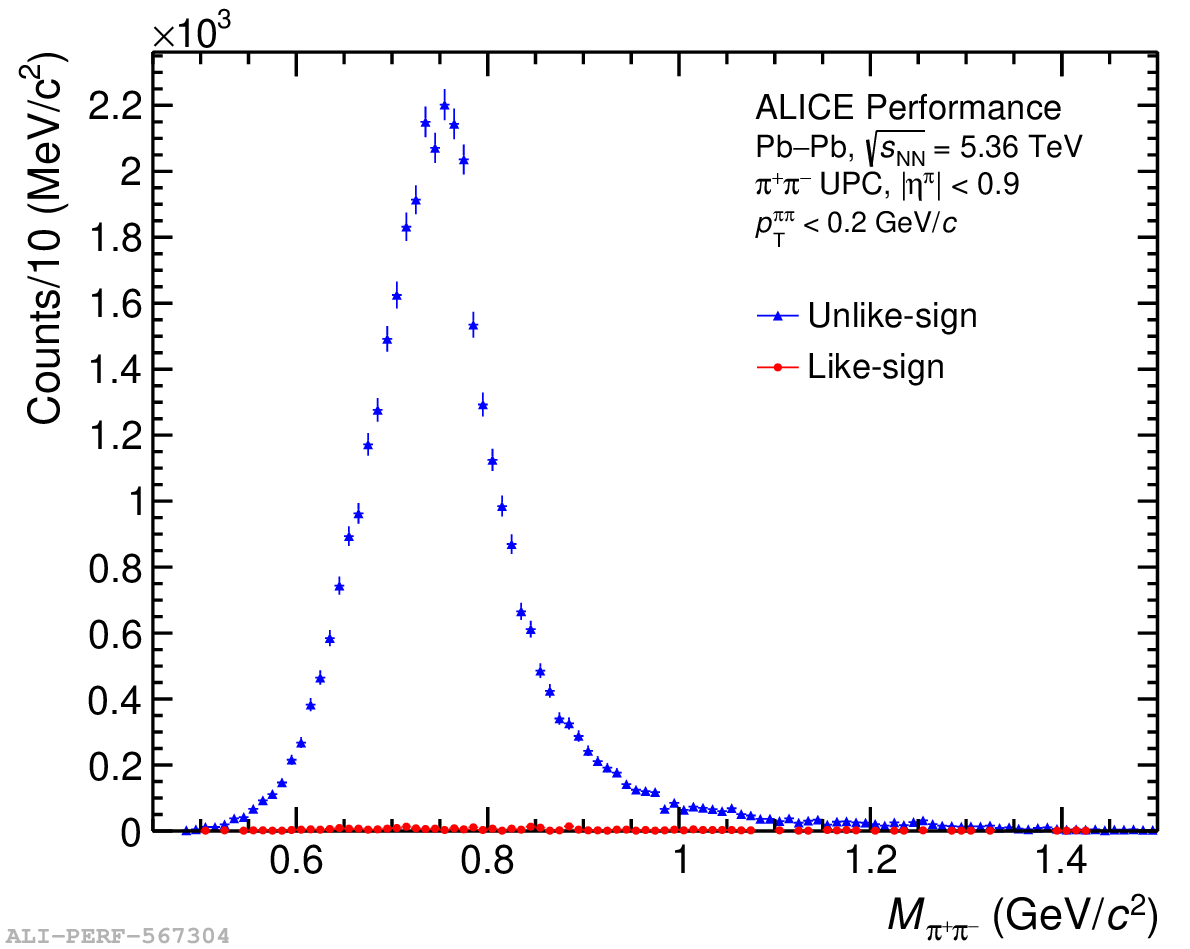}
    \includegraphics[width=0.48\textwidth]{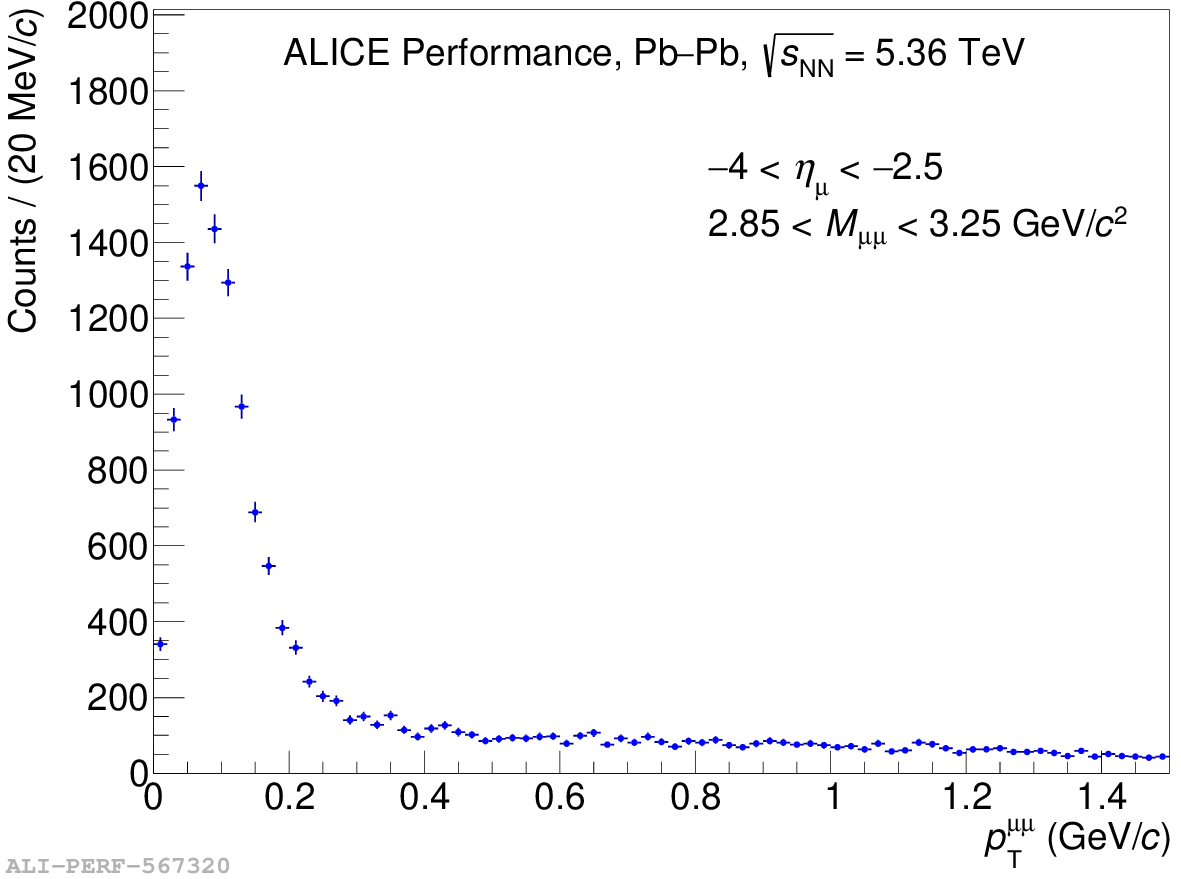}
    \caption{ Coherent $\rho^{0}$ invariant mass distribution (left) and \pt \ distribution of exclusive J/$\psi$ (right). }
    \label{invM}
\end{figure}

At present, among the ongoing photoproduction analyses using Run 3 data are the coherent $\rho^{0}$ at midrapidity and coherent J/$\psi$ at forward rapidity worth mentioning. These are chosen as Run 2 reference analyses to evaluate the performance of the ALICE reconstruction software in Run 3. The performance examples of these two analyses are shown in Fig.~\ref{invM}. Clear resonance peak of coherent $\rho^{0}$ in Pb-Pb UPCs at $\sqrt{s_{NN}}$ = 5.36 TeV is observed. The transverse momentum distribution of the photoproduced J/$\psi$ in Fig.~\ref{invM} shows both coherent and incoherent processes.

\section{UPC physics prospects in Run 3 and beyond}

\subsection{Exclusive vector meson photoproduction}
As discussed in Section~\ref{upceventrun3}, a significant increase in integrated luminosity allows the precision study of vector meson photoproduction in UPCs with Run 3 data. For example, experimental uncertainties for nuclear suppression factor for $\gamma + Pb \rightarrow \rm{J}/\psi + Pb$ processes are expected to be at the level of 4$\%$ \cite{yellowreport} while in Run 2 measurements these were of the order of $\sim$ 10 $\%$ \cite{nuclearsp}. It is also possible to analyse double vector meson photoproduction of higher quarkonium states such as $\psi(2S)$ and $\Upsilon(ns)$ with abundant exclusive UPC events~\cite{oops}. In addition, the Run 3 data opens the possibility of measuring strangeness in UPC in particular $\phi(1020) \rightarrow K^{+} + K^{-}$. Photo-production of $K^{+}K^{-}$ pairs has been measured with Run 2 data in ALICE \cite{kkpairs}. It will be possible to measure the exclusive photoproduction of a much wider range of particles, including excited vector mesons and searches for exotica, e.g. $X(3872)$.

\subsection{Exclusive vector meson photoproduction with FoCal}
\begin{figure}[h!]
    %\centering
    \includegraphics[width=0.49\textwidth]{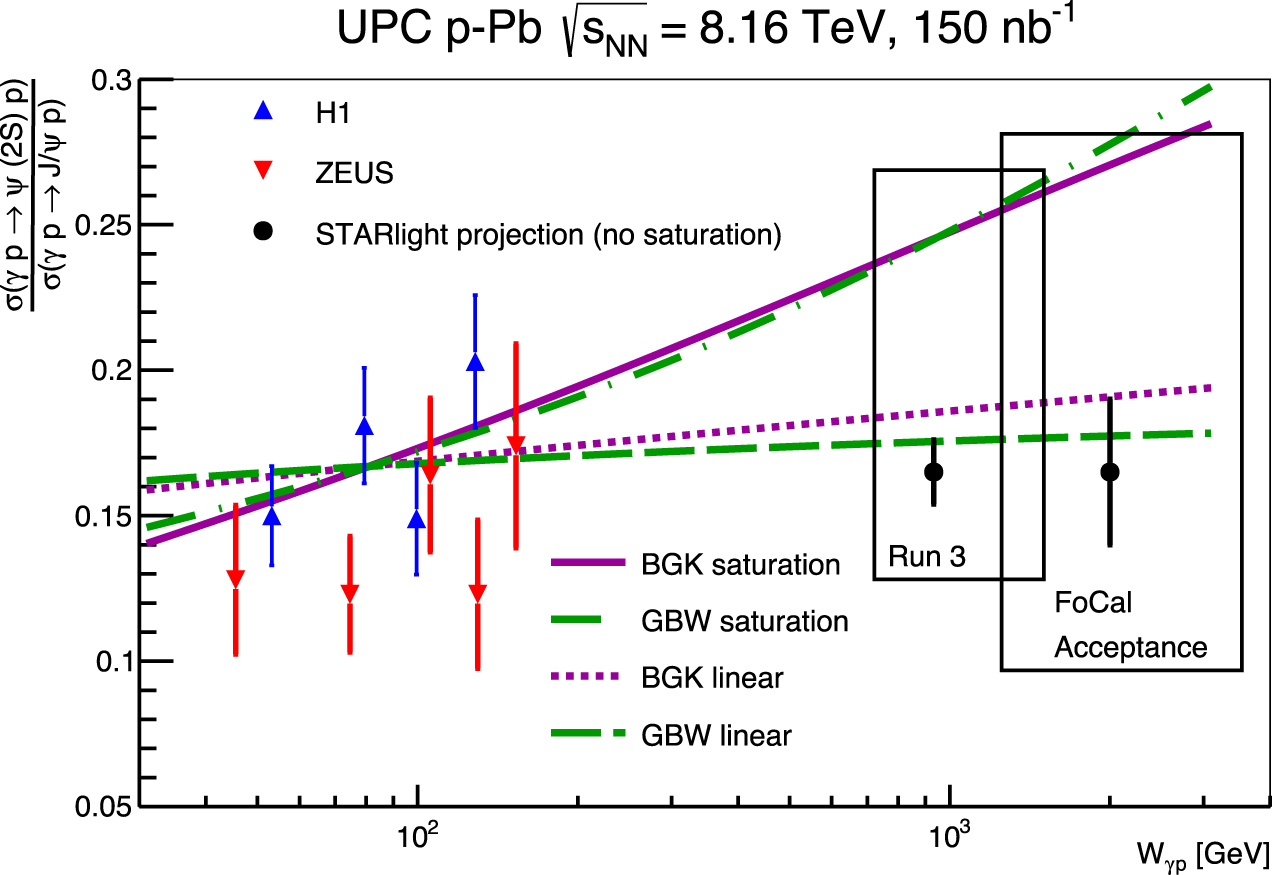}
    \includegraphics[width=0.49\textwidth]{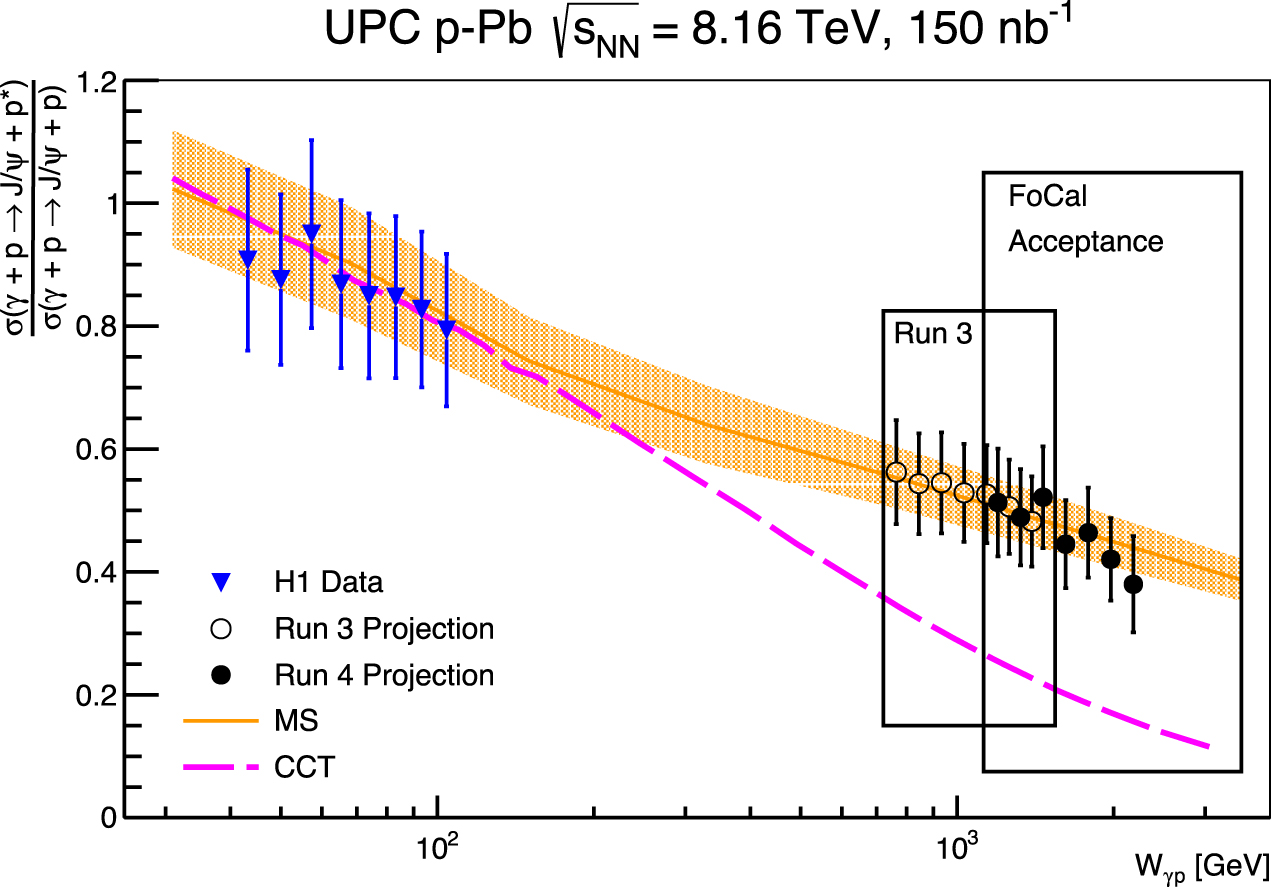}
    \caption{ Projection of ALICE Run 3 with Focal acceptance in Run 4 ratio of energy dependence: Exclusive $\psi(2S)$ to J/$\psi$ photoproduction cross section (left) and dissociative  J/$\psi$ to exclusive  J/$\psi$ photoproduction cross sections (right). }
    \label{focalprojection}
\end{figure}

The Forward Calorimeter (FoCal) is part of the ALICE upgrade for Run 4 (starting from 2029). It will be positioned 7 $m$ from the interaction point on the A-side, covering $3.4 < \rm \eta < 5.8$. As shown in Ref.~\cite{JDdissocuative}, the FoCal will provide access to the kinematic region where the gluon saturation phenomena dominate. The future measurements of the energy dependence of J/$\psi$ photoproduction will provide a direct model-independent probe for gluon saturation with large statistics and more forward acceptance in p-Pb and Pb-Pb UPC collisions. The ratio of exclusive $\psi(2S)$ to J/$\psi$ photoproduction cross section as a function of the energy ($W_{\gamma p}$) using the muon arm data in Run 3 alone shows sensitivity to the gluon saturation as shown in Fig.~\ref{focalprojection}. FoCal will enable a comprehensive study at the highest energy point (see Fig.~\ref{focalprojection}). The projections are also shown for dissociative J/$\psi$ in Run 3 with FOCAL acceptance in Run 4, which is sensitive to the gluon saturation \cite{JDdissocuative}.

\subsection{Search for new particles and inclusive UPCs}
The trigger-less data taking enables ALICE to study inclusive, inelastic photonuclear processes, e.g. inclusive J/$\psi$, jets in UPCs. The inclusive processes have never been studied at LHC so far; only results from HERA are available in the field of inelastic photonuclear interactions. This opens the possibility for studying open heavy flavours like $D^{0}$ in UPCs. Unlike vector meson photoproductions, a single gluon is involved, resulting in a 5 to 10 times larger cross section than for charmonia \cite{opencharm1, opencharm2}. 

\begin{figure}[h!]
    %\centering
    \includegraphics[width=0.49\textwidth]{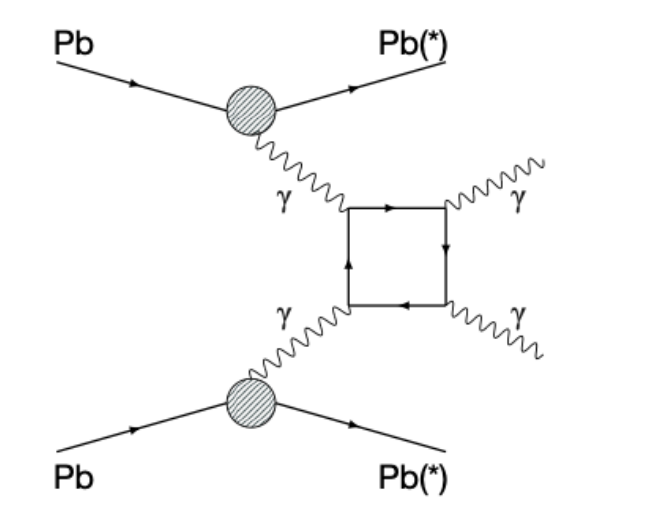}
    \includegraphics[width=0.49\textwidth]{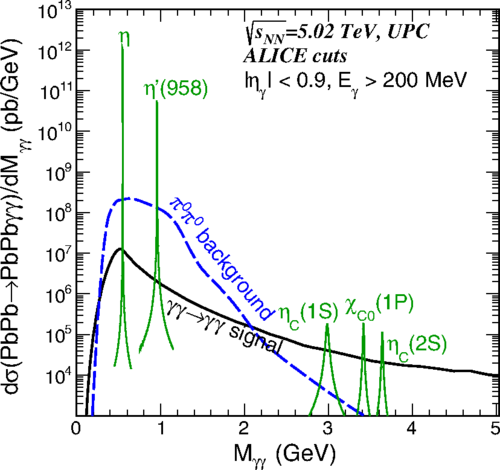}
    \caption{ The light-by-light scattering process (left) and contributions from resonances decays into two photons as a function of the invariant mass (right)~\cite{lyltheory}. }
    \label{LbL}
\end{figure}

The continuous readout also may enable the possibility to measure the rare fundamental QED process $\gamma\gamma \rightarrow \gamma \gamma$ interactions with ALICE, known as light-by-light scattering. The search for Axion Like Particles (ALPs) is done while measuring light-by-light scattering in Pb-Pb and looking for resonances in the invariant mass distributions as ALPs can couple to photons in the initial or final state of $\gamma\gamma \rightarrow \gamma \gamma$. Previously measured by ATLAS \cite{lybAtlas} and CMS \cite{lblCMS} in Run 2. As discussed here~\cite{lyltheory}, ALICE can potentially go down to 1 GeV to study elastic $\gamma\gamma \rightarrow \gamma \gamma$ scattering, focusing on low diphoton invariant masses below 5 GeV/$c^{2}$ (see Fig.~\ref{LbL}). However, the considered backgrounds (e.g. direct $\eta_{c}$ production) are underestimated as the calculation does not include soft photons coming from radiative decays of vector mesons such as $\gamma + \rm {Pomeron} -> \rm {J}/\psi -> \eta_{c}+\gamma$ \cite{SRklein}. Measuring ALPs is very challenging with the Run 3 setup as ALICE calorimeters can not measure low momenta neutral particles with high resolutions which will be possible in the Run 5 ALICE 3 program~\cite{alice3}.

\begin{figure}[h!]
    %\centering
    \includegraphics[width=0.47\textwidth]{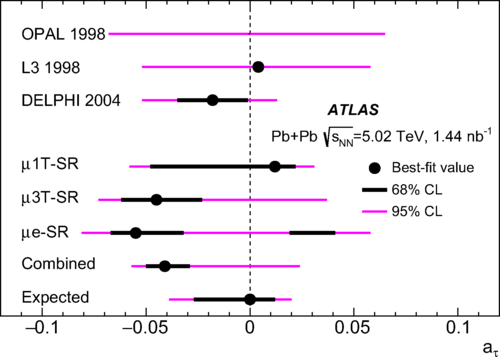}
    \includegraphics[width=0.50\textwidth]{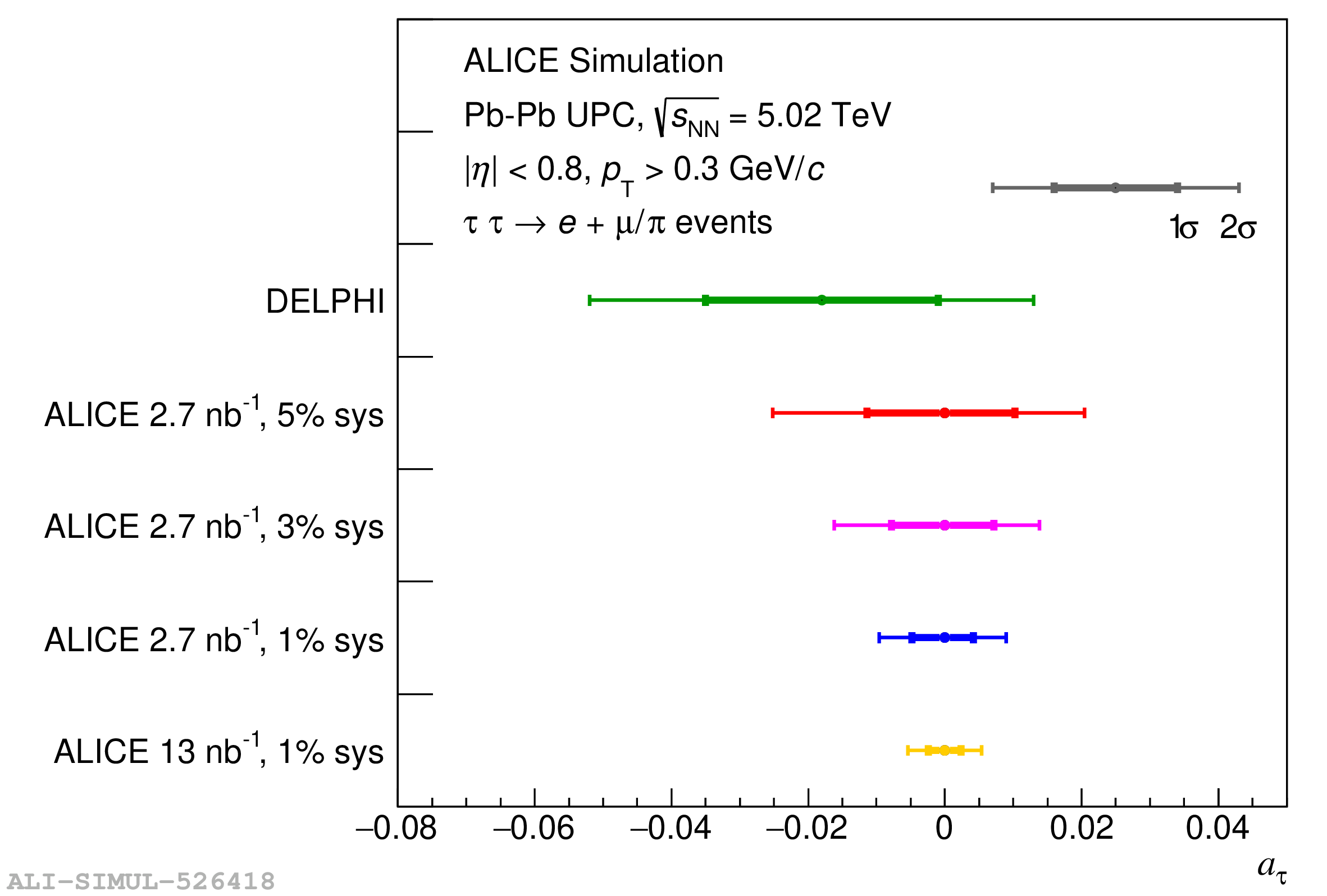}
    \caption{ The value of $a_{\tau}$ measured by ATLAS (left) and ALICE simulation with Run 3 acceptance for $a_{\tau}$ (right)~\cite{taoAtlas}. }
    \label{taoAM}
\end{figure}

Another potential scenario is the direct measurement of $\tau$ anomalous magnetic moment ($a_{\tau} = (g-2)/2$) in Run 3 with UPCs by studying $\gamma \gamma \rightarrow \tau^{+} \tau^{+}$ channel. The results from ATLAS ~\cite{taoAtlas} and CMS~\cite{taoCMS} using muon and one track ($e, \mu, hadron $), three tracks ( 3 hadrons) and electron topologies show precision similar to that achieved by DELPHI. This value is sensitive to many beyond-standard models (BSM) such as supersymmetry ($\delta a_{\tau} = m^{2}_{\tau}/M^{2}$). Run 3 provides luminosity to perform such studies in ALICE. The ALICE acceptance can allow us to perform these measurements down to low $p_{T}$ tracks and possibly constrains the uncertainty limits twice than is measured by ATLAS (Fig.~\ref{taoAM}).

In $\gamma \gamma$ interactions, it is also possible to study heavy tetra quark states $\gamma \gamma \rightarrow T_{4Q} \rightarrow 4l$ in UPC Pb-Pb collisions. Prediction for $X(6900) \rightarrow J/\psi J/\psi$ and $X(19000) \rightarrow \Upsilon \Upsilon$ within the ALICE kinematic range are discussed in Ref.~\cite{tetraquark}. An estimate of 100 events of charm tetra quark is expected before acceptance efficiency correction.

\section{Central exclusive production in pp collisions}
\label{cepsec}
\begin{figure}[h!]
    %\centering
    \includegraphics[width=0.97\textwidth]{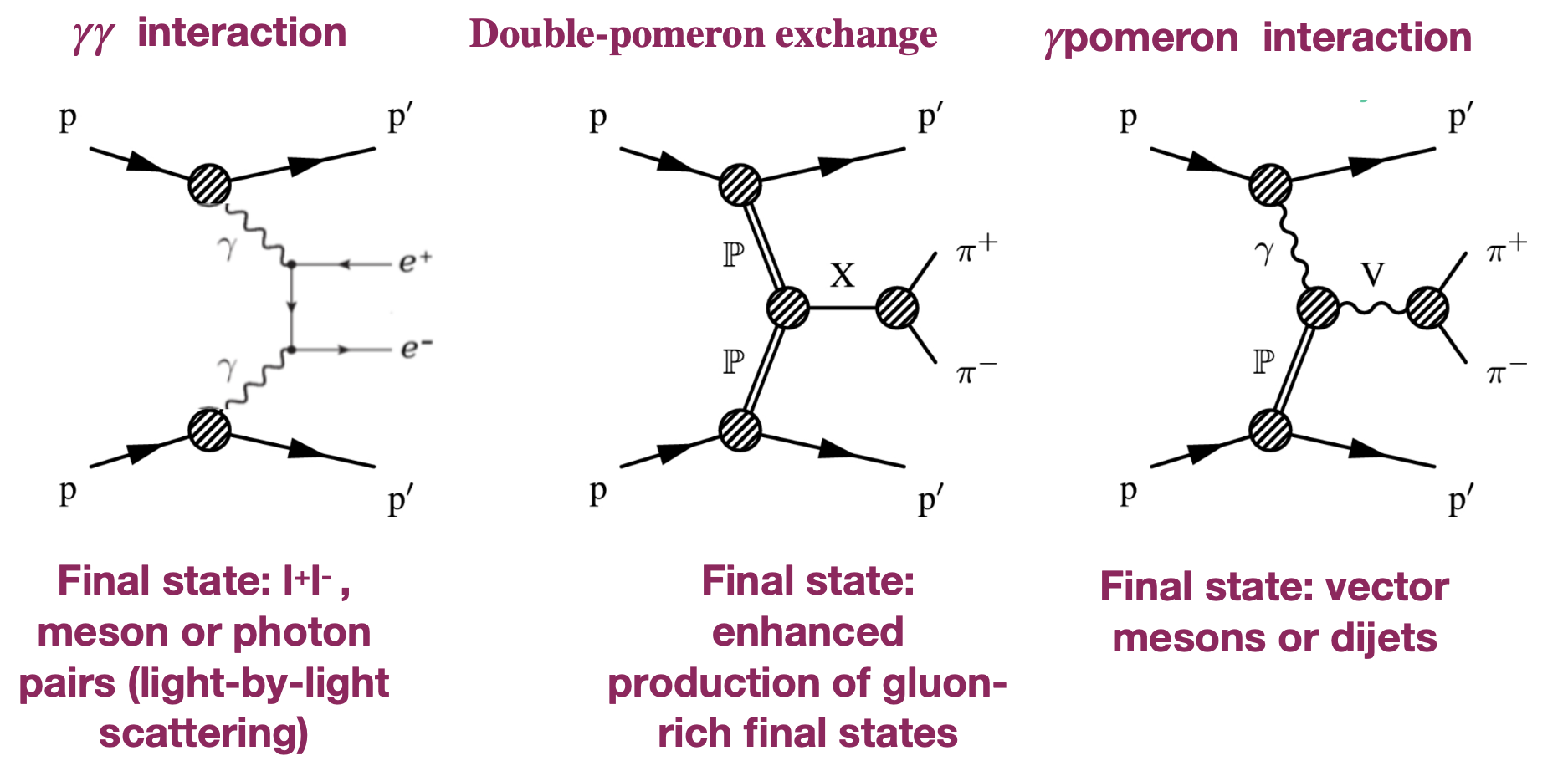}    
    \caption{ Schematic diagram of various CEP events. }
    \label{cepfig}
\end{figure}

Run 3 has opened the possibility to explore Central Exclusive Production (CEP) events in pp collisions to a new extent, which was not possible during Run 1 or Run 2. CEP events are studied using double-gap topology in the ALICE central barrel at mid-rapidity. Tracks are selected within the central barrel, having no signal on the FIT detectors. Several such diffraction processes are of interest, as shown in Fig.~\ref{cepfig}.

\begin{figure}[h!]
    %\centering
    \includegraphics[width=0.99\textwidth]{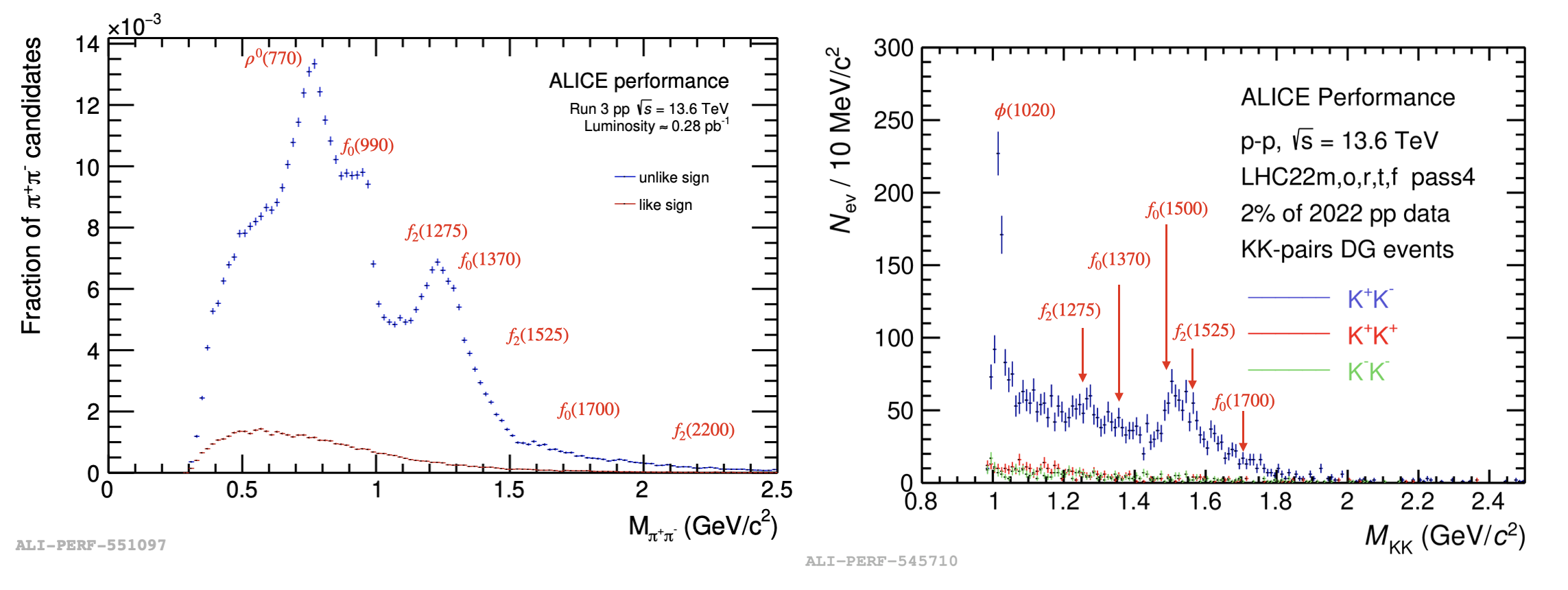}  \caption{Raw invariant mass distributions of two pions and two kaons candidates of CEP events in pp collisions.}
    \label{cepinVM}
\end{figure}

As mentioned in Section \ref{upceventrun3}, the minimum bias data collected in pp collisions in Run 3 in 2022 and 2023 is more than 3 times than Run 2. After using the double gap topology, the events are selected with two opposite charge tracks. Then, the particle identification is carried out by TPC down to low \pt \ based on specific energy loss (pion, kaon hypothesis). Such analyses show visible resonances in raw invariant mass distributions of opposite-sign pions and kaons (see Fig.~\ref{cepinVM}). It will be possible to study strangeness in double gap events with $\phi{1020}$ and $f_{2}(1525)$ states.

\section{Summary}
The field of UPCs in Run 3 and Run 4 is of great scientific interest. ALICE has already collected intriguing data from both Pb-Pb and pp collisions, with upcoming prospects for p-Pb data collection. These endeavours promise precision measurements of non-linear QCD effects such as saturation, the discovery of new resonances, and the exploration of new physics phenomena such as ALPs and Tetraquarks. UPCs also offer avenues for investigating strangeness, heavy-quarkonia like Upsilon states, and open charm. The new analysis framework is equipped to manage the anticipated event rates throughout Run 3 and beyond. 

\section*{Acknowledgments}
This research has been supported by the U.S. Department of Energy, Office of Science, Nuclear Physics.

\end{document}